\documentstyle[12pt]{article}

\begin{document}
\def\ket#1{\vert #1\rangle}
\def\be{\begin{equation}}
\def\ee{\end{equation}}
\def\ba{\begin{eqnarray}}
\def\ea{\end{eqnarray}}
\def\bZ{{\bf Z}}
\def\la{\langle}
\def\fa{\rangle}
\def\IZ{{\bf Z}}
\def\ket#1{\vert #1\rangle}
\begin{flushright}CLNS 93/1256\\October 1993\\hep-th/9311021
\end{flushright}
\bigskip
\centerline{\large {\bf STATUS OF FRACTIONAL SUPERSTRINGS}}

\bigskip\bigskip
\centerline{\large {\bf S.-H. Henry Tye}}
\centerline{\it Newman Laboratory of Nuclear Studies}
\centerline{\it Cornell University}
\centerline{\it Ithaca, NY  14853-5001}

\bigskip

\begin{abstract}

We argue why it is important to search beyond the superstring
for consistent string theories that have space-time supersymmetry
with critical space-time dimensions less than 10. We
give a lengthy introduction on the motivation and approach. We discuss
briefly some promising possibilities, namely, fractional
superstrings. More specifically, we consider the $K=4$, or spin-4/3
fractional superstring, which has a 3-dimensional flat Minkowski
space-time representation with bosons and fermions in its spectrum.
Its central charge of 5 is less than the critical value of 10.
The no-ghost theorem for the bosonic sector is proved, using the
recently constructed Kac determinants. We summarize the present
status and point out some open questions.
\\
\bigskip

Talk at the conference ``Strings 93'', Berkeley, May 23-29, 1993.

\end{abstract}

\section{Introduction}

	String theory is the only known theory with the potential
for describing all matter and forces in nature in a unified way.  In
particular, the superstring theory and the closely related
heterotic string theory entail many structures, including gravity,
gauge fields and chiral fermions, that are central to the
present understanding of our universe. However, their critical
space-time dimension is ten, and there
are numerous mechanisms to reduce the number of observable
space-time dimensions. Although the number of
models in ten dimensions is very limited,
namely the $SO(32)$ and the $E_8\otimes E_8$ models\cite{GHMR},
the number of models with four flat space-time dimensions is
huge. Furthermore, there is no known
compelling reason why the superstring
theory should have only four large
space-time dimensions.  While it is
important to search for dynamical
and/or symmetry reasons explaining how and why
our universe can be realized in the
heterotic/superstring framework, it is also important
to search for a new string theory that gives four large
space-time dimensions naturally. Whether such a string theory
exists or not is a question that we should be able to answer
definitively.

For self-consistency, it is reasonable to request that the new string
theory have space-time supersymmetry(SUSY). Let us recall
the superstring
case. ensional superstring model with spacetime
SUSY automatically has a zero cosmological
constant to all orders in perturbation theory.
On the other hand, a non-supersymmetric string model has, generically,
a non-zero cosmological constant through string loop corrections.
Such a cosmological constant is typically over a hundred orders of
magnitude too big. One may argue that
this does not happen in superstring theory, where space-time
SUSY is broken spontaneously.
The argument is based on two observations: (1) in string theory,
a non-zero cosmological constant will destroy the factorization
property in multi-loop string diagrams, thus invalidating
the consistency of string perturbation expansions; (2) starting with
a consistent quantum theory, spontaneous/dynamical symmetry
breaking will occur only in a self-consistent way. These observations
imply that, if a superstring model with spacetime
SUSY is a consistent quantum theory, then, after SUSY
breaking, it must remain a consistent theory, i.e., the
resulting cosmological constant in flat Minkowski spacetime
must remain precisely zero, since a consistent
string perturbative expansion must exist in the new
non-supersymmetric vacuum. Of course, this argument sheds no
light on whether or not (or how) SUSY breaking takes place
in string theory.

 So our goal is to search for the existence of new consistent
string theories that have space-time SUSY and
lower critical space-time dimensions.
Our approach starts from the string
world-sheet. String theories are characterized by the
local symmetries of two-dimensional conformal
field theories on the string world-sheet.
The bosonic string is invariant under diffeomorphisms
and local Weyl rescalings on the world-sheet;
whereas the superstring is characterized by
a locally supersymmetric version of these
symmetries. This involves a spin-$3/2$ supercurrent,
in addition to the energy-momentum tensor,
on the string world-sheet. It is natural to ask whether other
symmetries on the world-sheet can give rise
to consistent string theories, which have lower space-time dimensions.
To obtain a string theory from the world-sheet symmetry, the
symmetry algebra is treated as a constraint algebra on the string
Fock space. This approach have
been successfully applied to the bosonic string and the superstring.
It is natural to conjecture that, given any world-sheet symmetry algebra,
there is a corresponding string theory with that algebra as the constraint
algebra. Recently, a lot of progress have been made on the
W-string\cite{CP}, supporting this conjecture.

	The first step of this approach is to understand the properties of
world-sheet symmetries. I give an overall view of the
possible world-sheet symmetry algebras. Specific examples\cite{AGT} are
given to illustrate the richness of possibilities.
In Section 2, I discuss some specific possibilities of new
string theories, namely, fractional superstrings. Since the properties
of fractional superstrings were described in detail in the
literature[4-13],
I shall restrict myself to a brief summary of the present status and
some remarks. Then I shall discuss a specific example,
namely, the $K=4$, or spin-4/3 fractional superstring\cite{tree1}.
This theory is the simpliest non-trivial example, and is
characterized by a chiral algebra involving a pair of spin-4/3 currents
on the world-sheet in addition to the energy-momentum tensor\cite{ZFft}.
Scattering amplitudes of this theory are briefly
described. They satisfy both spurious state decoupling and  cyclic
symmetry (duality). Examples of such amplitudes are calculated using an
explicit $c=5$ realization of the spin-4/3 current algebra\cite{tree1}.
This representation has a 3-dimensional flat Minkowski space-time
interpretation, with fermions and bosons only in its
spectrum\cite{tree2}. The no-ghost theorem for the bosonic sector
is proved using the recently
constructed Kac determinants\cite{AKT}.
Again, I shall restrict myself to a brief introduction and a
summary of the present status. Section 3 contains some remarks.

	The structure of two-dimensional conformal field theories(CFT)
is in large part determined by their underlying conformal symmetry
algebra, which organizes all the fields in a
CFT into sets of primary fields of definite conformal dimensions,
and their associated infinite towers of descendant fields.  The
fundamental conformal symmetry is the Virasoro algebra.
The most general extended conformal symmetry algebra can be written
as follows. Consider a set of currents $J_i(z)$, primary with respect
to $T(z)$, with conformal dimensions $h_i$,
where $i\in\{1,2,\ldots,N\}$:
\ba\label{Iii}
  &&T(z)T(w)={{c/2}\over{(z-w)^4}}+{{2T(w)}\over{(z-w)^2}}+
  {{\partial T(w)}\over{(z-w)}}+...{}~,\\
 &&T(z)J_i(w)={{h_iJ_i(w)}\over{(z-w)^2}}
  +{{\partial J_i(w)}\over{(z-w)}}+.... \nonumber
\ea
The operator product expansions (OPEs) among the
currents have the following generic form:
\ba\label{Iiii}
  &&J_i(z)J_j(w)=~q_{ij}(z-w)^{-h_i-h_j}(1+.... )+\\
  &&\qquad\sum_kf_{ijk}(z-w)^{-h_i-h_j+h_k}
  \left(J_k(w)+.... \right).\nonumber
\ea
where $q_{ij}$ and $f_{ijk}$ are structure constants. The ellipses
stand for current algebra descendants,
whose dimensions differ from those of
the identity and the $J_k$ by positive integers.
Since the algebra is chiral, i.e.,
independent of $\bar z$, the conformal dimensions are the spins of
the fields. The parameters $c$, $h_i$, $q_{ij}$
and $f_{ijk}$ are not free and
must be chosen such that the algebra(\ref{Iiii}) is associative.
 This condition
places strong constraints on the set of consistent conformal dimensions
$h_i$ and restricts the structure constants $q_{ij}$ and $f_{ijk}$ as
functions of the central charge $c$.

Extended conformal algebras naturally fall into three classes:

(i) Local algebras: the simplest type, where all powers of $(z-w)$
appearing in (\ref{Iiii}) are integers. This class includes the most
familiar extended algebras, such as the superconformal,
Ka\v{c}-Moody and $W_n$
algebras.  These examples are unitary and therefore consist of currents
with only integer and half-integer spins.

When fractional powers of $(z-w)$ appear in some of the OPEs in
(\ref{Iiii}), some of the currents
will necessarily have fractional spins. In this case, the algebra is
non-local, due to the presence of Riemann cuts in the complex plane.
Such algebras are more complicated to construct and analyze than the
local ones.  Among non-local algebras there is a further division, again
along lines of complication.

(ii) Abelian non-local algebras: also known as parafermion (PF)
or generalized parafermion current algebras, were first constructed by
Zamolodchikov and Fateev\cite{ZFpf}.  They are the simplest type of
non-local algebras, involving at most one fractional power of $(z-w)$
in each OPE in (\ref{Iiii}).  Any two currents
in a PF algebra obey abelian braiding
relations, i.e., upon braiding the two currents
(analytically continuing one current along a path
encircling the other), any correlation function involving these two
currents only changes by a phase.  The analysis of the associativity
conditions for PF theories can be carried out using algebraic methods.

(iii) Non-abelian non-local algebras: or non-abelian algebras for short,
since they are necessarily non-local. This is the most general class of
extended algebras.  Their characteristic
feature is that their OPEs involve multiple cuts, i.e., there are
terms in at least one of the OPEs in (\ref{Iiii}) with different
fractional powers of $(z-w)$.
Any two fractional spin currents appearing in one of these OPEs will in
general obey non-abelian braiding properties.  The analysis of the
associativity conditions for non-abelian algebras requires more powerful
methods.

In general, the holomorphic $n$-point correlation functions of the
currents, $\la J_i(z_1)J_j(z_2).... J_k(z_n)\fa~$,
can be expressed
as a linear combination of some set of conformal blocks.  The relative
coefficients of the various conformal blocks are fixed by the closure
condition and the associativity condition.  The closure condition
is simply the requirement that no new currents beyond the currents of
the algebra should appear.  Associativity is the condition that the
particular linear combination of conformal blocks that appears in the
$n$-point function is invariant under fusion transformations (i.e.,
duality).  For the local algebras, each conformal block involves only
integer powers of $(z_i-z_j)$; for the abelian non-local algebras, each
correlation function of the parafermion currents has exactly one
conformal block, even though it involves fractional powers of $(z_i-z_j)$.
Of course, for the most general case we expect each correlation function
to have multiple conformal blocks, and the conformal blocks to involve
different fractional powers of $(z_i-z_j)$.  This general case corresponds
to the non-abelian non-local algebras. From this point of view we see
that upon braiding the currents, the correlation function is,
in general, transformed into an independent linear combination of
conformal
blocks.  This reflects the different phases that are picked up upon
analytically continuing the different fractional powers of $(z_i-z_j)$.

The different braiding properties of the different types of extended
algebras described above are reflected in the moding of their currents.
On a given state in any representation of the algebra, we can obtain new
states by acting with current modes
\ba\label{Iiv}
  J^{i_1}_{-n_1-r_1}J^{i_2}_{-n_2-r_2}... J^{i_m}_{-n_m-r_m}|\Phi\fa,
\ea
where the $n_j$ are integers and the $r_j$ are fractional in general.
For local unitary algebras, the half-integer spin currents can have
only integer or half-integer modings, and $r_1=r_2=\cdots=r_m=r$ where
 either $r=0$ or $r={1\over2}$, depending on the state $|\Phi\fa$. For
PF theories, the
situation is slightly more complicated.  Generically the $r_i$ are
different, determined by the state $|\Phi\fa$ and the currents that
preceded it. For non-abelian algebras, the moding of a particular
current in (\ref{Iiv}) is not unique; it depends both on the state
it operates on as well as on the state we want it to create.

Another useful way to classify symmetry algebras is their linearity
property. An algebra is said to be linear if tensoring any two of its
representations automatically gives a representation of the same
algebra. The Virasoro and the superconformal algebras are linear in
this sense. Most other algebras are non-linear, i.e., they do not
have this property. The construction of representations for these
algebras are rather non-trivial.

  The first set of non-abelian algebras was conjectured in
Ref.\cite{KMQ} and explicitly constructed in
Ref.\cite{AGT}. Consider a fractional supercurrent G with conformal
dimension (i.e., spin) $\Delta=(K+4)/(K+2)$. The following OPEs
define the level $K$ FSCA with arbitrary central charge $c$:
\begin{eqnarray}\label{FSCA}
 &&T(z)T(w)={(c/2)\over(z-w)^4}+{2T(w)\over(z-w)^2}+
   {\partial T(w)\over(z-w)}+...{}~,\nonumber\\
 &&T(z)G(w)={\Delta\over(z-w)^2}G(w)+
   {1\over(z-w)}\partial G(w)+...{}~,\\
 &&G(z)G(w)= {(z-w)^{-2\Delta}}\left\{{c\over\Delta}+
   2{(z-w)^2}T(w)+...\right\}+\nonumber\\
 &&\qquad\lambda(c){(z-w)^{-\Delta}}\left\{G(w)
 +{1\over2}(z-w)\partial G(w)+...\right\}.\nonumber
\end{eqnarray}
The associativity condition for this algebra fixes the structure
constant $\lambda^2 (c)$:
\begin{equation}\label{lambda(c)}
 \lambda^2 (c)={2{K^2}{c_{111}^2}\over{3(K+4)^2}}(c_1-c){}~.
\end{equation}
Here $c_{111}$ is the $SU(2)_K$ structure constant for the OPE of two
chiral spin-1 primary fields to give a chiral spin-1 field. It depends
only on $K$ and is explicitly given in Ref.\cite{AGT}; $c_1=24/K+c_0$,
and  $c_0$=$3K/(K+2)$.

For $K=1$, $G(z)$ is absent and we recover the Virasoro algebra.
For $K=2$, $c_{111}$
is zero and we recover the usual superconformal algebra, which
has no cuts in the compex plane in its OPEs. The absence of cuts is
an enormous simplification, and is the reason why the superconformal
algebra is substantially easier to study than the $K>2$ cases.
We call these algebras the fractional superconformal
algebras (FSCAs), since they are the natural generalization of the
superconformal algebra. The closed algebra generated
by the currents $T(z)$ and $G(z)$ is the level $K$ FSCA. It is
closely associated with the $SU(2)_K$ WZW model, in the sense that
a representation of level $K$ FSCA can be constructed from the currents
and the $j=1$ fields of $SU(2)_K$. We can also construct representations
using $\IZ_K$ parafermions.
We note that there exist other level $K$ FSCAs
with additional fractional supercurrents \cite{AGT}.
It is not hard to convince oneself that, for each affine Lie algebra,
there is at least one such corresponding extended conformal symmetry
algebra; some examples
are $W_N$ algebra associated with $SU(N)_1$\cite{ZFL} and the PF current
algebras associated with $so(N)_2$\cite{GoS}. The usefulness of these
symmetry algebras in
organizing CFTs is illustrated by the application of FSCA to
the $SU(2)$ WZW coset models\cite{KMQ,CLT}.

\section{Fractional superstrings}

	It is natural to ask whether symmetries other than the
superconformal symmetry on the world-sheet can give rise
to consistent string theories, which have lower space-time dimensions.
Recently, we find strong evidence for new string theories based on
FSCAs, called fractional
superstrings (FSS), which have interesting phenomenologies in
space-times with critical dimensions less
than 10. The evidence come from the following:

(1) Critical central charge and null states; any consistent string
theory is expected to have extra sets of physical null states at a
special value of the central charge, known as the critical
central charge. This property is presumably a reflection of the underlying
gauge symmetry of the string theory. Recently the Kac determinants for the
level $K$ FSCA have been constructed using the BRST cohomology techniques
developed in the conformal field theory\cite{AKT}. In particular, we
reproduce the Kac determinants for the Virasoro ($K=1$) and the
superconformal ($K=2$) algebras.
The Kac determinant formulae
allow us to examine the null state
structure of plausible critical FSS with an arbitrary level $K$
world-sheet fractional supersymmetry. The critical central charge is
found to be\cite{ALT,AKT}
\begin{equation}\label{critical}
 c_{\rm critical}=c_1+c_0={6K\over{K+2}}+{24\over K}{}~.
\end{equation}
Note that this recovers the well-known results for the bosonic ($K=1$,
with $c_{\rm critical}=26$) and the superstring ($K=2$, with
$c_{\rm critical}=15$) theories. This result also agrees with the more
explicit determination of the spin-4/3 ($K=4$, with
$c_{\rm critical}=10$) FSS given in Appendix E of Ref.\cite{ALyT}.

(2) One-loop modular invariant partition function; since the fractional
superstring involves world-sheet fractional spin fields, i.e.,
parafermions, we search for closed string modular invariant partition
functions that:
	(i)    involve  parafermion characters,
	(ii)   contain the massless graviton, and
	(iii)  are tachyon-free.
Rather unique solutions are obtained, and they are consistent with
space-time supersymmetry\cite{AT,ADT}. The FSS with spin-3/2 ($K=2$),
4/3 ($K=4$), 6/5 ($K=8$) and 10/9 ($K=16$) fractional supercurrent on
the world-sheet seem to have maximum flat dimension 10, 6, 4 and 3
respectively. Of course, the spin-3/2 case is simply the usual
superstring. It is natural to view the other
cases as non-trivial generalizations of the superstring case.

(3) Tree-level scattering amplitudes; we construct tree-level scattering
amplitudes for the spin-4/3 fractional superstring\cite{tree1}. Despite
the non-local and non-linear structure of the world-sheet symmetry,
we construct the N-point scattering amplitude for bosons and demonstrate
its cyclic symmetry (duality) and the decoupling of spurious states to
the physical states. In a particular representation which has a three
dimensional Minkowski space-time, we also argue that only space-time
bosons and fermions
are present in this model\cite{tree1,tree2}. The closed string version
 contains the massless graviton.

(4) No-ghost theorem; we apply the Kac determinant formula for
the K=4 FSC algebra to the above
three-dimensional Minkowski space-time representation of
the spin-4/3 fractional superstring\cite{AKT}. We prove the no-ghost
theorem for the space-time bosonic sector of this model; that is, at
the tree-level scattering, its
physical spectrum is free of negative-norm states.

Here, a few remarks on the above properties are in order:

(a) In the construction of the modular invariant partition functions,
we found the critical dimension to be, for $K>1$,
$D_{\rm crit}=2+{16/K}$.
There, we tensor ($D_{\rm crit}-2$) copies of $Z_K$ parafermions plus
boson $X^\mu$ in the light-cone gauge, each copy with central charge
$c_0$. Naively, this would imply that the critical central charge is
$c_{\rm crit}$=${D_{\rm crit}c_0}$=${6(K+8)/(K+2)}$,
which disagrees with (\ref{critical}). However, a remarkable
cancellation happens in the modular invariant partition functions,
within the characters of the sectors containing the massless fields.
This delicate cancellation, referred to as ``internal projection''
\cite{ADT}, reduces the central charge in the light-cone gauge from
$c_{\rm lc}=(D_{\rm crit}-2)c_0=48/(K+2)$ to an effective value
$c_{\rm lc}=24/K$. This can agree with (\ref{critical}) if we add
back the central charge $2c_0$ for the time and longitudinal
dimensions\cite{AD}.
This ``internal projection'' feature is not surprising, for the
following reason. The level $K$ FSCA are non-linear for $K>2$, so its
representations cannot be obtained from tensoring copies of $Z_K$
parafermion plus boson $X^\mu$. Inside the space of tensored
copies, one can imagine a projection to a subspace of states
which does form a representation of the algebra. The explicit
construction discussed later in this section is motivated by this
observation. Since the $K=2$ FSCA
(the superconformal algebra) is linear,
there is no internal projection for the superstring case.

(b) It was argued that the critical dimensions for the $K=4$ and
$K=8$ FSS are 6 and 4 respectively. However, a closer
examination\cite{DT} of the modular invariant partition functions
for these two cases indicates that they do not have the necessary
Lorentz symmetry:
the $K=4$ modular invariant partition function allows at most
a four-dimensional Poincare symmetry while the $K=8$ case allows
at most a three-dimensional Poincare symmetry. This seems to imply
that the two of the flat directions in the $K=4$ FSS cannot be normal
space-time dimensions. To avoid a continuous spectrum, they must be
compactified. This maximum possible space-time dimension was
referred to as the ``natural dimension'', as opposed to the critical
dimension. In this sense, the $K=4$ FSS is the phenomenologically
interesting theory to pursue, since its natural space-time
dimension is 4. As we shall see in the next section,
the $K=4$ FSS is also by far the easiest case (among all FSS
with non-linear FSCA) to study.

(c) In uncompactified superstrings, the spectrum does not contain
chiral fermions in fundamental representations of gauge groups.
 They appear only after appropriate compactification.
On the other hand, it is the compactification that removes the
uniqueness of the superstring theory, and raises the non-trivial
dynamical questions of why and how does it happen. So it is intriguing
to see that the $K=4$ FSS has natural dimension of 4, which is
less than its critical dimension of 6; in this situation, one can
hope the compactification
of the extra dimensions to appear more naturally\cite{DT}. Of course,
an explicit realization of this feature remains to be found.

	To be concrete, let us consider the $K=4$ FSCA. This is the
simpliest non-trivial case because one can split the $G$ current
into two pieces, $G(z)=G^+(z)+G^-(z)$,
which then satisfy abelian braid relations:
\ba\label{ftalg}
  &&G^{\pm}(z)G^{\mp}(w)={1\over(z-w)^{8/3}}\left\{{3c\over8}+
   (z-w)^2T(w)+...\right\},\\
  &&G^{\pm}(z)G^{\pm}(w)={\lambda\over(z-w)^{4/3}}
   \left\{G^{\mp}(w)+{1\over2}(z-w)\partial
   G^{\mp}(w)+...\right\}.\nonumber
\ea
Here, ${\lambda^2}=(8-c)/6$. In this split form, this algebra
is the spin-4/3 para-fermionic current algebra constructed by
Zamolodchikov and Fateev \cite{ZFft}. Appendix C
of Ref.\cite{ALyT} gives a detailed construction of this algebra.
To construct a string theory based on this algebra, we use an approach
that mimics the original construction of the superstring\cite{RNS}.
By analogy with the superconformal gauge of the superstring, the
stress-energy tensor and fractional supercurrents are assumed to
generate the physical state conditions.  In particular, physical
states, taken to be annihilated by all the positive modes $L_n$ and
$G^\pm_r$ of $T$ and $G^\pm$ respectively,
are the highest-weight states of the FSC algebra. This FSC algebra
has a $\IZ_3$ symmetry.
In particular, the currents $G^{+}$ and $G^{-}$
can be assigned $\IZ_3$ charges $q=1$ and $-1$, respectively, while
the energy-momentum tensor $T$ (as well as the identity) has charge
$q=0$. So the highest-weight modules of this algebra
are organized by the $\IZ_3$ symmetry of the
algebra.  Highest-weight states with
$\IZ_3$ charge $\pm1$ are said to belong to D-modules, while those
with $\IZ_3$ charge 0 are in S-modules.
The consistency of the FSS can be demonstrated
by defining tree scattering amplitudes and showing that
they are consistent with the assumed physical state conditions
following from the spin-4/3 FSCA.  In
other words, in tree scattering, physical states never scatter to
unphysical states, and null states can also be consistently decoupled
from scattering of other physical states.  This property is commonly
referred to as {\it spurious state decoupling}, and can be shown
in a representation-independent way.  The argument for
spurious state decoupling follows closely that used in the ``old
covariant formalism'' \cite{NST} for ordinary superstring amplitudes;

Scattering of D-module states can be written in three physically
equivalent ``pictures,'' reflecting the $\IZ_3$ symmetry of the
fractional superconformal algebra, in which the vertex operators
for scattering can be one of $W^\pm$ of conformal dimension 1/3
and $\IZ_3$ charge $\pm1$ or V of conformal dimension 1
and $\IZ_3$ charge 0. In the old covariant formalism, the
N-point scattering amplitude can be written in three pictures
\ba\label{Apict}
        {\cal A}_N &=& 2\langle W^+_N| V_{N-1}(1)
        \Delta ....\Delta V_2(1) \ket{W^-_1} \nonumber\\
        &=& 2\langle W^-_N| V_{N-1}(1)
        \Delta ....\Delta V_2(1) \ket{W^+_1} \nonumber\\
        &=& \langle V_N| V_{N-1}(1)\widetilde\Delta
        ....\widetilde\Delta V_2(1) \ket{V_1} .
\ea
Here the propagators are
$\Delta=(L_0-{1\over3})^{-1}$ and $\widetilde\Delta=(L_0-1)^{-1}$;
and the vertex operators are related by
 \be\label{gocom}
        [G^\pm_r,V(1)]=\left(L_0+r-{1\over3}\right)W^\pm(1)
        -W^\pm(1)\left(L_0-{1\over3}\right)
 \ee
 for all $r\in\IZ/3$.
This is closely analogous to the two different pictures for scattering
of Neveu-Schwarz sector states in the
old covariant formalism for the ordinary superstring, in which vertex
operators can be either G-parity even dimension-1/2 operators or
G-parity odd dimension-1 operators (this property follows from
the $\IZ_2$ symmetry of the superconformal algebra). Scattering of
S-module states is problematic due to the absence of an
appropriate dimension-1 vertex operator in that sector.
Since the S-module typically contains the tachyonic state,
we are fortunate that they can be decoupled
from tree scattering amplitudes of D-module states by a $\IZ_3$
analog of the GSO projection \cite{GSO}.
A separate issue that can be addressed at tree level is the
unitarity of scattering amplitudes.  In particular, spurious state
decoupling implies unitarity only if one can prove that the space of
physical states has non-negative norm.  This latter property is called
the {\it no-ghost theorem}.
Using the recently constructed Kac determinants for FSCAs,
such a theorem is proven in Ref.\cite{AKT} for
the three-dimensional model that we shall now discuss.

This model is a particular conformal field theory
representation of the spin-4/3 FSC algebra with
central charge $c=5$.  It is made up of three free coordinate boson
fields $X^\mu$ on the world-sheet and a two-boson representation of
the $so(2,1)_2$ WZW model, which is closely related to
the $SU(3)_1$ WZW model; (note that, conformally,
$SU(3)_1$=$SU(2)_4$=$so(3)_2$). This model has a global
three-dimensional Poincare invariance.  The non-linear nature of
the spin-4/3 FSC algebra makes the existence of
such a representation non-trivial.  Also, the states in the model are
found to be space-time bosons or fermions, showing that the existence
of fractional-spin constraints on the world-sheet need not imply
fractional spins in space-time. The untwisted sectors of this
FSCA describe space-time bosonic physical states in
this representation. In particular, the lowest-mass D-module states
describe massless gauge fields for the open string and a graviton
for closed fractional superstrings.

 Since three-dimensional Minkowski space-time is too small to
decribe nature, it is encouraging that the central charge of the
three-dimensional model is less than the critical value of 10,
allowing the possibility of a critical spin-4/3 fractional
superstring containing four-dimensional Minkowski space-time.

\section{Discussion and outlook}

	Of course, the existence of FSS remains to be
demonstrated. There are many open questions remain to be answered.
More specifically, we are still steps away from showing that

(1) consistent fractional superstring theory exists and

(2) it is phenomenologically interesting.

However, we are encouraged by some non-trivial ``coincidences''
that are worth emphasizing. Let me mention two:

(1) In the construction of modular invariant partition functions,
typically one finds there are more constraints than the number
of parameters available in finding a solution. A priori, solutions
should not exist, but they do.

(2) In the construction of scattering amplitudes for D-module
states in the $K=4$ FSS, which includes the massless vector
particle as the ground state, the constraints necessary for spurious
state decoupling and for picture-changing to work seems to be tighter
than the freedom allowed by the parameters present. Naively, the
picture-changing formula
(\ref{gocom}) should not exist a priori, but it does.

We believe that these ``coincidences'' are really a reflection of an
underlying consistent theory being looked at from a funny angle.
They clearly suggest that FSS is tighter than the usual
superstring theory. This can be seen in other ways as well:

(1) the number of possible modular invariant partition functions
of both the FSS type and the heterotic type are a lot fewer than in
the corresponding cases in usual superstring, at any fixed space-time
dimensions\cite{DT}. This comes about because of (i) the initial critical
dimension of FSS starts out to be smaller than that of the superstring;
so there are less dimensions to compactify; and (ii) the FSCAs underlying
the FSS are non-linear, hence there are fewer representations for fixed
central charge.

(2) the tachyonic state (or any state in the same sector) in the
Neveu-Schwarz sector is completely
consistent in the superstring theory, even though it is undesirable
phenomenologically; the same tachyonic state (or any state in the
same sector) in the $K=4$ FSS is not consistent: the four-point
scattering amplitude of such states does not exist in the old
covariant formalism. So, the GSO-like
projection is necessary in the FSS case even at the tree-level.

At the moment, we are trying to construct the BRST ghosts
and the BRST invariance of the $K=4$ FSS. Here, the ghosts
are expected to couple to the matter fields, much like the Yang-Mills
theory. In this case, one should seek a $c=0$ representation of the
FSC algebra that contains both the matter and the ghost fields and
permits the construction of a nilpotent BRST charge.
Within this $c=0$ representation, the goal is to find a $c=10$
sub-representation that has a four-dimensional Poincare invariance.
Unfortunately, there is no clear guideline on how to search for such a
representation. It is interesting to note
that, since $so(4)=so(3)\otimes so(3)$, the world-sheet symmetry algebra
corresponding to $so(4)_2$ is simply two copies of the spin-4/3
algebra; however, the coordinate bosons coupled to the $so(3)_2$ model
in our $c=5$ representation do not transform in the vector
representation of $so(4)$, and so cannot give a flat space-time
interpretation.

In general, other fractional superstrings will be technically
more difficult to work with than the spin-4/3 FSS.
The main reason is that the general FSC algebra are non-abelian.
Since conformally, $su(2)_{2L}=so(3)_L$, we expect all of them (at
least the ones with even $K$) to
have a representation (similar to the $c=5$ one in the $K=4$ FSS)
with a three-dimensional Poincare invariance. However, like the
$K=4$ case, none of them will be at the critical central charge
except when $K$ is infinite. In this limit, the fractional supercurrent
has conformal dimension 1, with $c_{\rm critical}=6$ for the
corresponding FSS. The representation
is made up of three free coordinate boson fields $X^\mu$ on the
world-sheet and a three-boson representation of the
$so(2,1)_{\infty}$ WZW model, so the central charge equals
$c_{\rm critical}$. This string theory has critical
space-time dimension 3, and it may be used to
describe the three-dimensional Ising model. Since the  OPEs in
this $K=\infty$ FSCA has only
poles, the analysis may be quite tractable.

\begin{flushleft}
\large\bf Acknowledgements
\end{flushleft}

This talk reports on joint work with Stephen Chung, Keith Dienes,
Jim Grochocinski, Zurab Kakushadze, Andre LeClair, Ed Lyman, and
especially Philip Argyres. It is a pleasure to thank them all
for fruitful collaborations and many useful discussions. This work
was supported in part by the National Science Foundation.

\end{document}